\documentclass[12pt,preprint]{aastex}

\slugcomment{submitted to ApJ}

\shorttitle{Abundance anticorrelations}
\shortauthors{Salaris et al.}

\begin{document}


\title{On the primordial scenario for abundance variations 
within globular clusters. The isochrone test.}


\author{Maurizio Salaris\altaffilmark{1}}
\affil{Astrophysics Research Institute, Liverpool John Moores
University, 12 Quays House, Birkenhead, CH41 1LD, UK}
\affil{Max Planck Institut f\"ur Astrophysik,
Karl-Schwarzschild-Strasse~1, Garching, D-85748}
\email{ms@astro.livjm.ac.uk}

\author{Achim Weiss}
\affil{Max Planck Institut f\"ur Astrophysik,
Karl-Schwarzschild-Strasse~1, Garching, D-85748}
\email{weiss@MPA-Garching.MPG.DE}

\author{Jason W. Ferguson}
\affil{Physics Department, Wichita State University, Wichita, KS 67260-0032}
\email{jason.ferguson@wichita.edu}

\and

\author{David J. Fusilier}
\affil{Physics Department, Wichita State University, Wichita, KS 67260-0032}
\email{djfusilier@wichita.edu}


\altaffiltext{1}{Guest User, Canadian Astronomy Data Centre, which is
  operated by the Dominion Astrophysical Observatory for the National
  Research Council 
of Canada's Herzberg Institute of Astrophysics.}

\begin{abstract}
Self-enrichment processes occurring in the early stages of a 
globular cluster lifetime are generally invoked to explain the
observed CNONaMgAl abundance anticorrelations within 
individual Galactic globulars.
We have tested, with fully consistent stellar evolution calculations,  
if theoretical isochrones for stars born with the observed abundance
anticorrelations satisfy the observational evidence that 
objects with different degrees of these anomalies lie on essentially
identical sequences in the Color-Magnitude-Diagram (CMD). To this purpose, 
we have computed for the first time low-mass stellar models and
isochrones with an initial metal mixture that includes the extreme
values of the observed abundance anticorrelations, 
and varying initial He mass fractions. 
Comparisons 
with 'normal' $\alpha$-enhanced isochrones and suitable Monte Carlo
simulations that include photometric errors show that a significant  
broadening of the CMD sequences occurs only if the helium 
enhancement is extremely large (in this study, when $Y$=0.35) in the
stars showing anomalous abundances.
Stellar luminosity functions up to the Red Giant Branch tip are also
very weakly affected, apart from -- depending on the He content of the 
polluting material --  the Red Giant Branch bump region.  
We also study the distribution of stars
along the Zero Age Horizontal Branch, and derive general constraints 
on the relative location of objects with and without abundance
anomalies along the observed horizontal branches of globular clusters.
\end{abstract}

\keywords{Hertzsprung-Russell diagram -- 
globular clusters: general --- stars: abundances -- stars: evolution --
stars: horizontal-branch}

\section{Introduction}

It is more than thirty years ago that CN variations 
in the atmospheres of stars within a single globular cluster were
first detected (Osborn~1971). Following this discovery, signs of 
hetereogeneity in the Na (Cohen~1978), Al (Norris et al.~1981)
and O (Leep, Wallerstein \& Oke~1986) abundances were also found.
Thanks to the large body of spectroscopic data published in the
intervening years, it is by now well established that surface 
abundance variations of
C, N, O, Na, and often -- but not always, see, e.g., Ramirez \&
Cohen~(2003) -- also of Mg and Al, exist in stars within individual
globular clusters (see, e.g., the recent review by Gratton, Sneden \& Carretta~2004).
There is now convincing evidence (e.g., Gratton et al.~2001, Carretta
et al.~2005, Cohen \& Mel\'endez~2005)  
that these elements display a pattern of abundance variations that is 
very constant all along the Red Giant Branch 
(RGB) down to the Turn-Off (TO) region. The pattern shows anticorrelations
between CN and ONa (and, when observed, MgAl) 
overimposed onto a 'normal' $\alpha$-enhanced 
heavy element distribution ([$\alpha$/Fe]$\sim 0.3-0.4$) 
in the sense that negative variations of C and O 
are accompanied by increased N and Na abundances.
%
%
The generally accepted hypothesis is that the presently existing stars were born with the
observed CNONa abundance patterns. 
Intermediate-mass Asymptotic Giant Branch (AGB) stars have often been 
invoked as sources of the necessary heavy element pollution (see, e.g. Ventura et
al.~2001, Carretta et et al.~2005 and references
therein). The envelopes of these AGB stars are 
expected to show a pattern of CNONa anticorrelations
similar to the observed one, and are expelled after a time of the
order of $\approx10^8$ yr during the  
thermal pulse phase. Provided that a significant fraction of the material is not
lost from the globular cluster, new stars can form directly out of this matter,
or from pre-existing matter polluted to varying degrees by the AGB
ejecta. The iron abundances would be constant among the different 
subpopulations, as is actually observed (see, e.g., Suntzeff~1993). 

If this scenario is correct, the
observed abundance anomalies contain valuable information about the
early stages of globular cluster formation and evolution.
Current AGB stellar models are not yet able to give a
definitive quantitative prediction of the element 
abundances in the envelopes of intermediate-mass AGB stars due
to a number of theoretical uncertainties (see, e.g., Ventura \& D'Antona~2005), 
but it is well established that matter polluted by AGB
stars has to show an He enrichment with respect to the original
cluster chemical composition. 
An alternative scenario invokes pollution from mass lost from evolved RGB stars 
which experienced extra-deep mixing 
(Denissenkov \& Weiss~2004). The expelled matter would not be 
He-enhanced in this case.

In this paper we assess if theoretical stellar isochrones
representative of stars born out of matter showing the observed abundance
anomalies, are compatible with the narrow
sequences traced by the observed broadband Color-Magnitude-Diagrams (CMDs) of
most of the Galactic globular clusters (GCs) -- where stars with
different degrees of these anticorrelations lie on essentially
identical sequences -- and their luminosity functions. 
This test has not been performed in a fully consistent manner until now. 
Only D'Antona et al.~(2002) have addressed the issue, but they 
investigated the effect of an increased 
helium abundance only at constant $Z$. In our analysis we have computed 
models, isochrones and luminosity functions of stellar populations 
born with both a normal  
$\alpha$-enhanced heavy element mixture and, for the first time, 
with a metal distribution displaying typical extreme values of 
the observed CNONa anticorrelations. As a further step we also
investigate if the theoretical interpretation of the observed CMDs and
star counts can
help us to discriminate between the AGB- and RGB-pollution scenarios.     
In the next section we briefly decribe the models, while Section~3
will analyze the results. Conclusions, and considerations about the
effect on the models of also the MgAl anticorrelation  
follow in the last section.

\section{The models}

We have computed sets of stellar models, theoretical isochrones and
Zero Age HB (ZAHB) sequences for [Fe/H]=$-$1.6, a
representative GC metallicity, appropriate for 
clusters like M3, M13 and NGC~6752 (according to the 
Zinn \& West~1984 [Fe/H] scale) which all display the CNONa
abundance anomalies (see, e.g., 
Carretta et al.~2001 for NGC6752, Sneden et al.~2004, Cohen \&
Melend\'ez~2005, Johnson et al.~2005 for M3 and M13). 
We have employed the same stellar evolution code as in 
Salaris \& Weiss~(1998) and the same sources for input physics, with
exception of the low temperature opacities, for which we rely 
on the code and database of Ferguson et al.~(2005) instead of 
Alexander \& Ferguson~(1994). 
The 'reference' metal abundance distribution is the $\alpha$-enhanced
one by Salaris \& Weiss~(1998), with $<$[$\alpha$/Fe]$>$=0.4. 
Our reference
models and isochrones employ this heavy element mixture, $Y$=0.25
(appropriate for GCs, see e.g. Salaris et al.~2004) and $Z$=0.001. 
We have then considered a second metal mixture, with typical mean
abundances of the extreme CNONa anticorrelations, 
based on the results summarized in Carretta et
al.~(2005). Assuming that both the CN and NaO anticorrelations are present 
at the same time in a given star (see, e.g., Table~1 in Carretta et
al.~2005) we have
modified the reference $\alpha$-enhanced mixture by an 1.8~dex increase
of N, an 0.6~dex decrease of C, an 0.8~dex increase of Na, 
and an 0.8~dex decrease of O. 
To fully take into account the 
effect of this modified or 'extreme' mixture in our models,  
stellar opacity tables have been specifically computed for this investigation. For 
temperatures higher than 10000~K we used the tabulations produced with 
the OPAL online facility at
\url{http://www-phys.llnl.gov/V Div/OPAL/new.html}, that are
consistent with their counterpart for the reference mixture.   
For lower temperatures, appropriate tables were computed 
with the code presented by Ferguson et al.~(2005).
Models and isochrones account for this
extreme mixture in both the nuclear network and opacities (the effect
of metals on the equation of state in the GC metallicity range is
negligible, therefore we used the same tables in both our sets of
models). They have $Z$=0.0018, and two different values of the initial He
abundance, $Y$=0.25 and 0.29 respectively. This latter value is taken
as representative of the enhanced He abundance in the matter processed by
contaminating AGB stars (see D'Antona et al.~2002). 
The value of $Z$ has been chosen to obtain the same [Fe/H] as in the
reference case (changing $Y$ from
0.25 to 0.29 at fixed $Z$ changes the corresponding [Fe/H] by only
0.02~dex so that no adjustment of $Z$ for taking into account the
enhanced $Y$ is necessary). As the Fe abundance is kept equal to the reference
mixture, the sum (C+N+O) is about 0.3~dex larger by number, 
a value consistent with the results by Carretta et al.~(2005) for the extreme values of the
anticorrelations. 

As a third case that will be discussed separately, we also explored the
possibility of a much larger $Y$ value for the extreme
mixture, by computing a set of models, isochrones and ZAHBs with $Y$=0.35.

Luminosities and effective temperatures of the isochrones for the
two metal mixtures have been transformed into magnitudes and colors using
in both cases the $\alpha$-enhanced color transformations by Cassisi et
al.~(2004) for [Fe/H]=$-$1.6. These transformations are in principle not
adequate for the extreme mixture, however, if comparisons are
performed in the $V-(V-I)$ plane, the inconsistency is
minimized because $(V-I)-T_\mathrm{eff}$
transformations are to a good approximation independent of the
metals and their distribution (e.g. Alonso et al.~1996, 1999, Cassisi et al.~2004).
We have verified this point by applying to our extreme isochrones and
ZAHBs color transformations corresponding to a metallicity increased by a
factor of 3 (a factor larger than the difference
in the global metal content between the reference and extreme
isochrones) with respect to the reference value discussed above. The
net result is that the $M_V$ of both isochrones and ZAHBs is shifted
by the same -- negligible -- amount of 0.017~mag. The $(V-I)$ colors
are unchanged.   

\section{Results and discussion}

\begin{figure}
\epsscale{.80}
\plotone{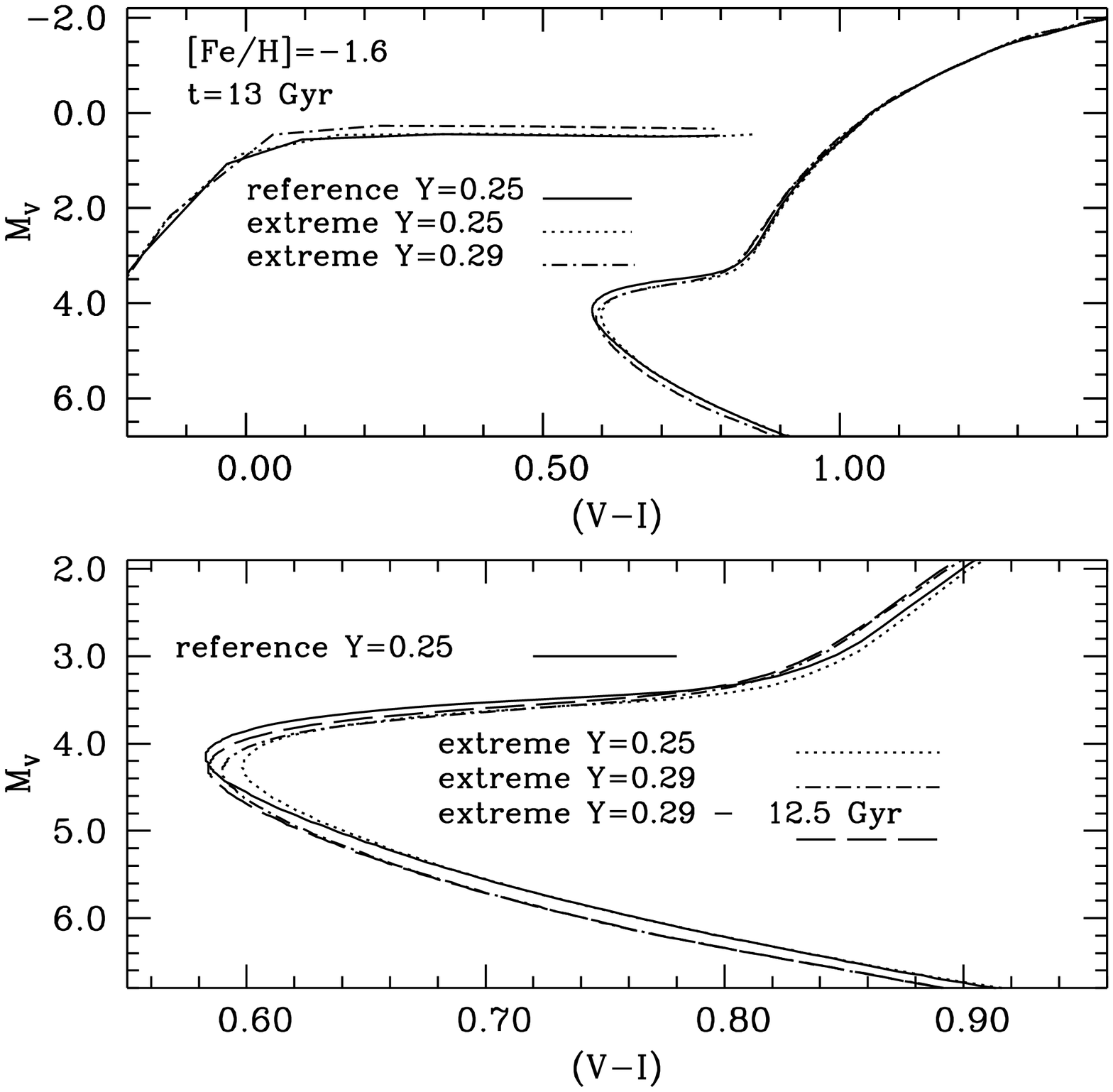}
\caption{CMD of 13~Gyr old isochrones with the 'reference'
  and 'extreme' metal mixtures, [Fe/H]=$-1.6$ and the labelled values
  of $Y$.
  \label{fig1}}
\end{figure}

The upper panel of Fig.~\ref{fig1} compares a set of 
13~Gyr old, [Fe/H]=$-$1.6 isochrones from the MS to the
tip of the RGB, plus the corresponding ZAHB sequence, for the three
different chemical compositions. Notice the complete overlap of
the RGB sequences for the reference mixture and the two isochrones
with the extreme CNONa anticorrelation and differing intial He mass
fractions. This fully satisfies the strong constraint posed by the very
narrow RGB sequences in the observed CMDs.  
Although the He core mass at the He flash is slightly lower for the
extreme mixture with $Y$=0.25 (0.483$M_{\odot}$ vs.\ 0.490$M_{\odot}$ for the
reference case; the envelope He mass fraction after the first
dredge-up is $Y$=0.262 in both cases) the ZAHB level is about 0.02~mag brighter. 
This is due to the higher value of the sum (C+N+O) in the extreme mixture.
The extreme mixture with $Y$=0.29 produces a
ZAHB about 0.16~mag brighter than the $Y$=0.25 case.  
Along the MS the $Y$=0.29 extreme isochrone differs in color by
only $\sim$0.02~mag from the reference one. 
The extreme mixture with $Y$=0.25 instead produces a MS in complete agreement with
the reference case. 
The lower panel of Fig.~\ref{fig1} enlarges the MS, TO and Subgiant
Branch (SGB)
region of the isochrones, and displays an additional line
corresponding to the extreme mixture, $Y$=0.29 and an age of
12.5~Gyr. Thereby we consider a slightly younger age for the anomalous subpopulation.
Results analogous to what is described in the following
are obtained even if the age of this subpopulation is closer to
13~Gyr. The younger age obviously has no 
effect on the RGB and ZAHB location.
If AGB stars are the main polluters, the 
most appropriate counterpart of a subpopulation
showing the extreme CNONa anticorrelation is likely to be the 12.5~Gyr old
isochrone with $Y$=0.29. In this case the TO has the same color of the
reference case, and it is underluminous by $\sim$0.15 mag. 

To check if these small differences along the MS and TO are compatible
with the observed CMDs, we made the following experiment: Using a Monte
Carlo algorithm we have populated both the 13~Gyr old reference isochrone and
the $Y$=0.29, 12.5~Gyr old one with the extreme mixture,  
using a Salpeter~(1955) Initial Mass Function and the same number of
objects. The individual magnitudes and colors have been then perturbed by a 
Gaussian (extremely small) 1$\sigma$
photometric error of 0.005~mag in both $V$ and $I$.  
This photometric error is equal to the mean error in the extremely 
accurate $VI$ photometries of a large sample of GCs by Stetson~(2000).
\footnote{
Data can be retrieved from \url{http://cadcwww.hia.nrc.ca/standards}}.

\begin{figure}
\epsscale{.80}
\plotone{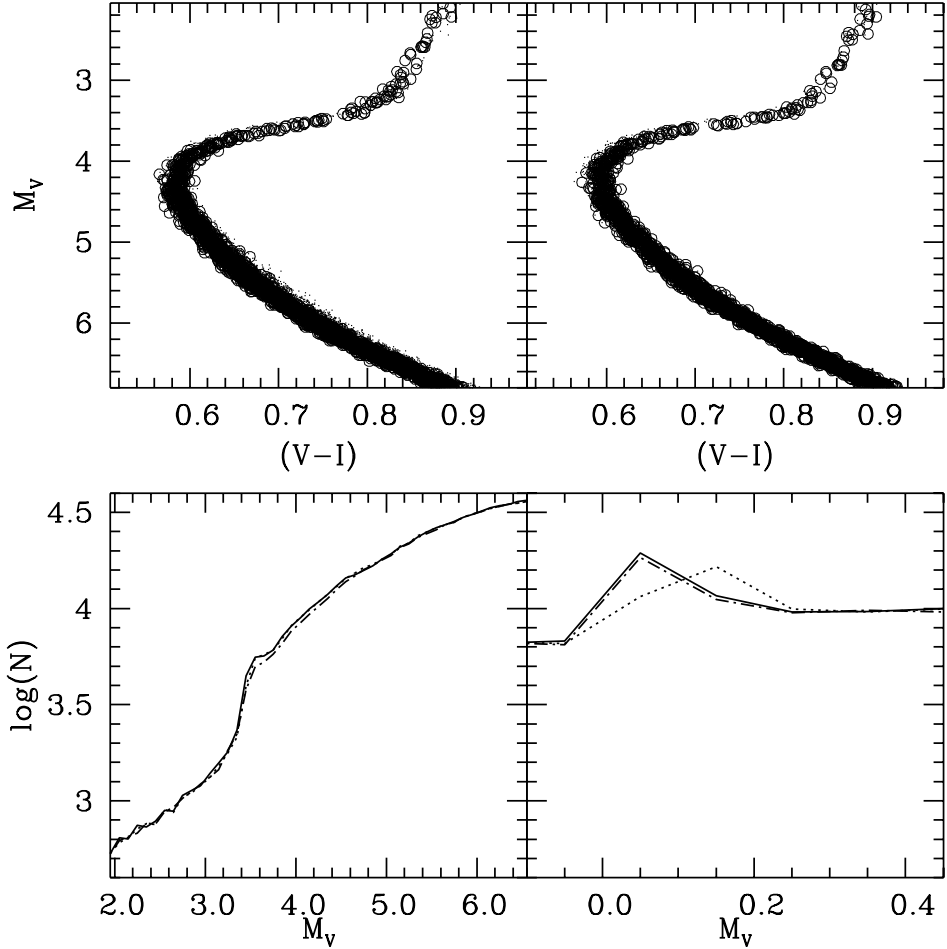}
\caption{CMD of a synthetic GC with [Fe/H]=$-$1.6. 50\% of the cluster
  stars formed out of the reference mixture and are 13~Gyr old (dots); the
  remaining 50\% formed out of the extreme mixture (open circles) 
  with either $Y$=0.29 (left upper panel) or $Y$=0.25 (right upper panel) and  
  have an age of 12.5~Gyr. A 1$\sigma$ Gaussian photometric error of
  0.005~mag is included in the simulation. The lower panels
  compare the luminosity functions of these two composite populations (dotted line
  for the case with $Y$=0.25, dash-dotted line for the case with $Y$=0.29) 
  with the case of a synthetic cluster made only of stars 
  formed out of the reference mixture (solid line) for the MS, TO, 
  base of the RGB (left lower panel) and the RGB bump region (left right
  panel). The star counts are   
  normalized to the same number of stars along the lower MS, with an
  arbitrary zero point that is different in the two panels.  
\label{fig2}}
\end{figure}

The result of the simulation is displayed in Fig.~\ref{fig2}.
Even with such a small photometric error, stars with the
reference and extreme mixture show a large overlap along the MS and
are completely coincident in the TO region and along the RGB (the
upper RGB is not displayed since already from Fig.~\ref{fig1} one can
notice that reference and extreme isochrones with both $Y$=0.25 and 
$Y$=0.29 overlap completely).
One has also to take into account that this is a somewhat extreme
case, given that stars in real clusters 
show a range of compositions (and maybe ages?) between these two extreme cases.

The lower panels of Fig.~\ref{fig2} show how the luminosity function (with 
magnitude bin of 0.10~mag, typical of the best observational counterparts) 
obtained from the combined 50\% reference+ 50\% extreme population, 
compares to the case of a cluster made of stars all 
formed out of the reference mixture. The two luminosity 
functions are essentially identical in shape, apart
from minute differences in the TO region.
The location and shape of the RGB bump is unchanged. 

In case the chemical composition of the polluting matter has 
$Y$=0.25, Fig.~\ref{fig2} shows 
the same Monte Carlo simulation as discussed for the case with $Y$=0.29
(the age for isochrone with the extreme composition is again 12.5~Gyr). Apart from
the identical MS, ZAHB and RGB that can be seen in Fig.~\ref{fig1},  
one can notice a largely overlapping TO and SGB region. 
The luminosity function of the 50\% reference+ 50\% extreme population
is again essentially identical to the case of a cluster
made of stars all formed with the reference mixture, apart from a
small shift (by about 0.10~mag)  
of the center of the RGB bump region towards fainter $M_V$. 

The behavior of the bump brightness in Fig.~\ref{fig2} deserves a
brief discussion. It is well known that at fixed age, the bump becomes fainter when
the metallicity of the isochrone increases at constant helium, whereas
it gets brighter when the initial helium abundance increases at
fixed metallicity (see, e.g., Salaris \& Cassisi~2005, Cassisi \&
Salaris~1997, Caloi \& D'Antona~2005). 
Also, the bump feature has a 
non-negligible intrinsic width, of the order of 0.2~mag. 
The center of the bump region in the extreme population with $Y$=0.25
is fainter than the reference one by $\sim$0.10~mag, due to its higher
total metallicity $Z$ (the small age
difference between reference and extreme population does not play a
major role). In a composite population made of stars with both reference and
extreme compositions, the two bump regions overlap partially, due to their
intrinsic widths, and their convolution produces a single
well defined feature, as in real GCs (e.g. Zoccali et al.~1999). 
Different weights of the reference and extreme components will
produce different shifts of the magnitude of the composite bump, compared to the
reference case.
When $Y$=0.29 for the extreme mixture, the bump center is fainter by
only 0.02~mag compared to the reference one, due to the increase of
its luminosity produced by the higher initial $Y$. The composite
population will therefore show a bump at essentially the same brightness of 
the reference isochrone.
In general, once the $Y$ value of the extreme population is fixed, 
both the center and the shape of the resulting composite bump feature 
are modulated by the relative weights of the two components.

\subsection{The case of an extreme mixture with $Y$=0.35}

\begin{figure}
\epsscale{.80}
\plotone{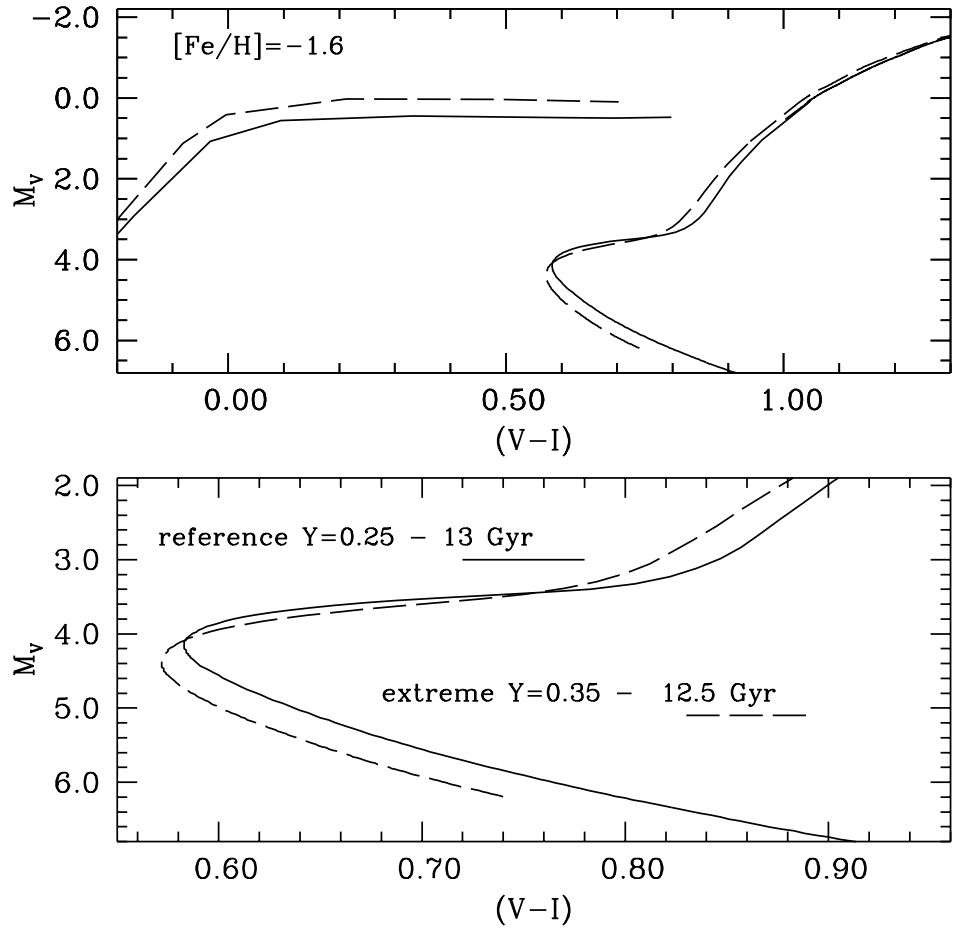}
\caption{As in Fig.~\ref{fig1}, but for the labelled mixtures. The
fainter end of the MS in the $Y$=0.35 isochrone corresponds to a mass
of 0.55$M_{\odot}$. 
\label{fig3}}
\end{figure}

Figure~\ref{fig3} compares the   
13~Gyr old reference isochrone and ZAHB, with a 12.5~Gyr
old counterpart computed with the extreme mixture and $Y$=0.35.
The differences found for the case of $Y$=0.29 are, as expected, 
amplified. The MS below the TO is bluer on average by $\sim$0.05~mag in the
extreme isochrone, its TO fainter by 0.25~mag (and bluer by
0.01~mag) the lower RGB bluer by $\sim$0.03~mag, 
this latter difference reducing almost to zero in the upper part of the RGB.
The ZAHB is brighter by $\sim$0.45~mag compared to the reference mixture.

\begin{figure}
\epsscale{.80}
\plotone{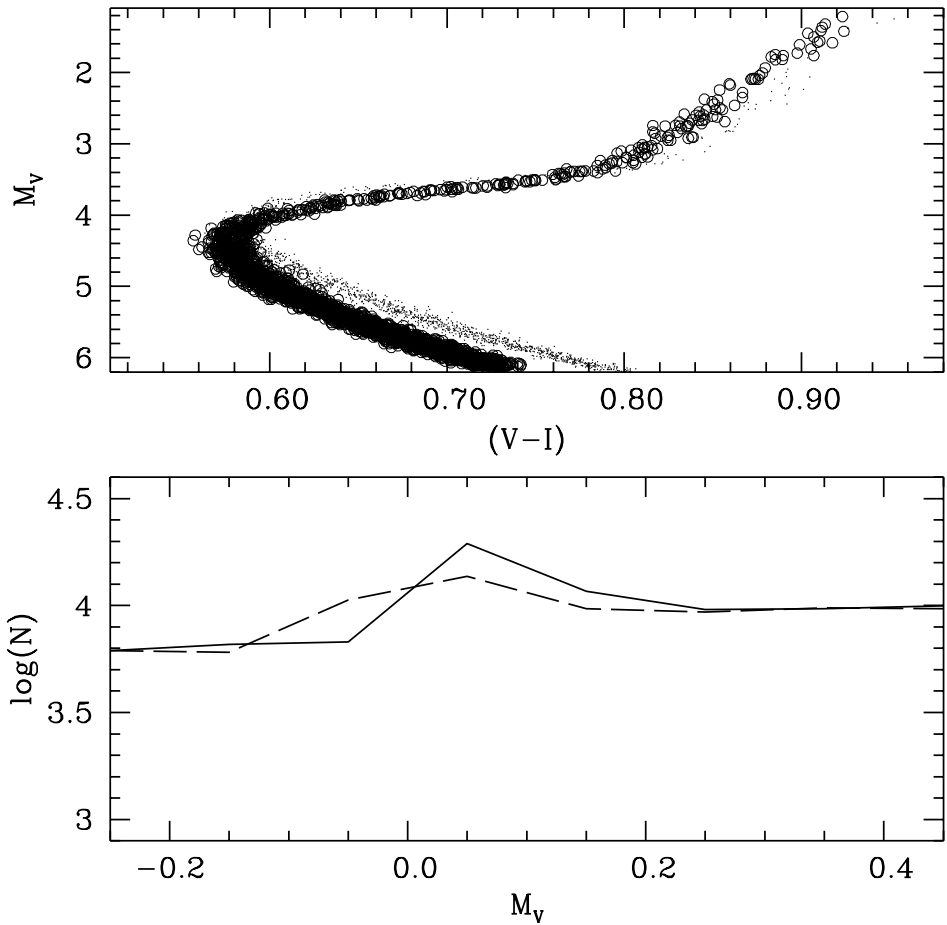}
\caption{CMD of a synthetic GC with [Fe/H]=$-$1.6 (upper panel). 50\% of the cluster
stars formed out of the reference mixture and are 13~Gyr old (dots); the
remaining 50\% formed out of the extreme mixture (open circles) 
with $Y$=0.35 and have an age of 12.5~Gyr. A 1$\sigma$ Gaussian photometric error of
0.005~mag is included in the simulation. The lower panel
compare the RGB bump region of luminosity function of this composite
population (dashed line) with the case of a synthetic cluster made only of stars 
formed out of the reference mixture (solid line). The star counts are   
normalized to the same number of stars along the lower MS. 
\label{fig4}}
\end{figure}

Figure~\ref{fig4} displays a synthetic CMD of a composite population
similar to the case discussed above.
50\% of the cluster stars formed out of the reference mixture and are 13~Gyr old (dots); the
remaining 50\% formed out of the extreme mixture (open circles) 
with $Y$=0.35 and have an age of 12.5~Gyr. A 1$\sigma$ Gaussian photometric error of
0.005~mag is included. 
The MS of the two subpopulations are now clearly separated in the
CMD, whereas the TO and SGB regions have a large degree of
overlap. The lower RGB of the $Y$=0.35 subpopulation is also 
separated from the reference one.
We have verified that 1$\sigma$ photometric errors of $\sim$0.03~mag
in $V$ and $I$ would merge the two MS sequences into a single one. 
An accurate analysis of the width of the observed MS of individual
clusters is necessary to conclusively assess if a component with 
such a large initial He abundance is present. 
It is perhaps interesting to notice that subcomponents with very 
high initial $Y$ have been discovered in one cluster, 
namely NGC2808 (see, eg., Carretta, Bragaglia \& Cacciari~2004 for
a recent spectroscopic study of the abundance anomalies in this
cluster) in addition to the well known case of
$\omega$~Centauri, that shows however also a spread in [Fe/H]  
(see, e.g., Suntzeff \& Kraft~1996
Norris~2004, Sollima et al.~2005, Piotto et al.~2005 and references
therein) not existing in NGC~2808 (e.g. Carretta et al.~2004). 
D'Antona et al.~(2005) found indications of a super He-rich
($Y\sim$0.40) population from an analysis of 
the width of its MS.

Figure~\ref{fig4} displays also the RGB bump region of the luminosity
function of the composite population described before, compared to the
case of a reference isochrone. The center of the bump is
displaced towards brighter luminosities by no more than
$\sim$0.05~mag. The two bump features show a large overlap and a
somewhat different shape. These differences are due to the fact
that although the $Y$=0.35 extreme isochrone produces a bump that is about
0.15~mag brighter than the reference isochrone (due to the much 
larger initial He abundance) it is also less pronounced (because of
the less efficient first dredge up at this high $Y$, and smaller discontinuity of the
H-profile) and still overlaps partially with the bump of the reference
population. The convolution of the two bump regions with the 50:50 weights
produces this composite feature.
The rest of the luminosity function shows only minute differences
(comparable to the case of the composite mixtures displayed in
Fig.~\ref{fig2}) around the TO region. 

\subsection{Populations along the Horizontal Branch}

A major difference between populations formed out of the reference
and extreme mixtures is the ZAHB brightness at colors redder than 
$(V-I)\sim$0.0 (see Fig.~\ref{fig1}). At bluer colors the ZAHBs become
more vertical and tend to overlap. 
A theoretical prediction of the expected color of the ZAHB (and the
whole HB) in the reference and extreme
populations is however difficult, given the lack of theoretical
understanding of the mass loss processes along the RGB phase. 
What is generally done is to assume an average amount of mass lost
along the RGB (for
example using the Reimers~1975 mass loss law and choosing appropriately its free
parameter $\eta$) and a spread around this value, that best reproduce the observed color
extension of the HB in individual clusters.
Here we investigate in the most general way how models with reference and
extreme mixtures (both present within clusters showing the
abundance anomalies) can coexist along the HB sequence, 
irrespective of the actual HB morphology in individual clusters. 

\begin{figure}
\epsscale{.80}
\plotone{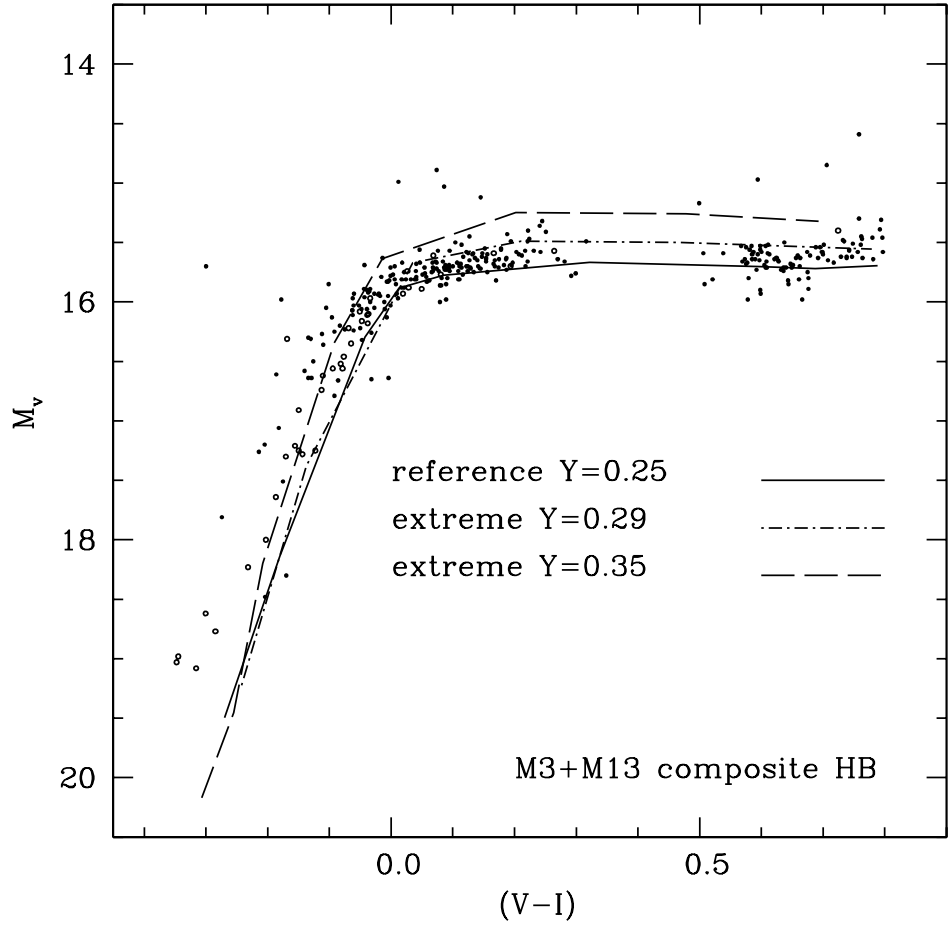}
\caption{CMD of ZAHBs for the labelled chemical compositions, compared
to a composite HB obtained combining observations M3 (filled circles)
and M13 (open circles -- see text for details). 
\label{fig5}}
\end{figure}

Figure~\ref{fig5} displays the CMD of non-variable HB stars in the Galactic GCs M3
and M13 (data from Ferraro et al.~1997 and Stetson~2000,
respectively) that both show the abundance anticorrelations discussed
in this paper. The HB of M13 has been first shifted in color to account for
its slightly different reddening compared to M3 ($E(B-V)$=0.01 for M3
and $E(B-V)$=0.02 for M13, following Dutra \& Bica~2000) and then in $V$-magnitude until the two
HBs overlap along the common blue part. In this way, one obtains an
empirical template at the [Fe/H] of the models, that covers
approximately the whole possible HB
color extension (the gap located between $(V-I)\sim$0.3 and $\sim$0.5 corresponds to
RR Lyrae instability strip) and poses as an empirical constraint to
the models. 
The ZAHB for the reference mixture is also plotted, shifted in color
to account for the reddening of M3, and in $V$, until the approximate lower
envelope of the observed HB is matched (given that the ZAHB marks the
start of the HB phase). To this end we followed the procedure
described in Salaris \& Weiss~(1997). We have looked
into the brightness-distribution of HB stars in a few color bins
along the horizontal part of the observed HB. For
each colour bin count histograms for brightness bins were created; 
the brightness bins were typically 0.05~mag wide (depending on
the HB population). We set the ZAHB level to the upper brightness of that bin
which shows a decrease in star  counts by a factor $\geq$2 and where the brighter bins contain
more than 90\% of all candidate HB stars under consideration. 
The ZAHBs for the 
extreme mixtures with $Y$=0.29 and $Y$=0.35 (the case of $Y$=0.25
is not shown, given that its ZAHB is essentially identical to the
reference one, as already discussed and shown in Fig.~\ref{fig1}) are
also displayed, after we applied the same color and magnitude shifts
employed for the reference ZAHB. 

We can now study two different possibilities, in the reasonable assumption
that the ZAHB of stars with the reference mixture coincides with the
observed one. 
We consider first the case of an RGB mass loss that populates the
three ZAHBs in the horizontal part, where 
$(V-I)\geq$0. With this assumption, one can easily notice that the
$Y$=0.29 models are 
located almost at the upper envelope of the empirical HB, and the
$Y$=0.35 one at even brighter magnitudes. This choice for the RGB mass
loss of the extreme populations is probably still 
viable for the $Y$=0.29 case, but only if the percentage of stars showing
the extreme anticorrelation is negligible. In fact, in this case only a
few stars of the entire HB population will be
evolved above the $Y$=0.29 ZAHB, therefore not contradicting the
evidence of very few objects present above this sequence. The $Y$=0.35 case appears to
be definitely ruled out, because the ZAHB is essentially above the upper
limit of the observed data distribution. The case of an extreme mixture with $Y$=0.25 is still
viable, given that the corresponding ZAHB is essentially coincident
with the reference one.
It is only in the perhaps unrealistic case 
that the fraction of stars with the reference chemical composition 
is negligible, that the extreme populations with
either $Y$=0.35 or $Y$=0.29 can also be present in this HB color range.
In fact, in these conditions it would be the ZAHB of the extreme populations  
that should be matched to the observed counterpart, by applying an
additional shift towards fainter magnitudes.

As a second case we consider an RGB mass loss that populates 
the ZAHBs in the color range $(V-I)<$0.0. 
In this case the reference ZAHB and the extreme one with $Y$=0.29 
(the one with $Y$=0.25 as well) overlap
completely, and can both coexist without contradiction with the empirical data. 
When $(V-I)$ is below $\sim -$0.15 also the $Y$=0.35 ZAHB tends to
overlap with the others.

\begin{figure}
\epsscale{.80}
\plotone{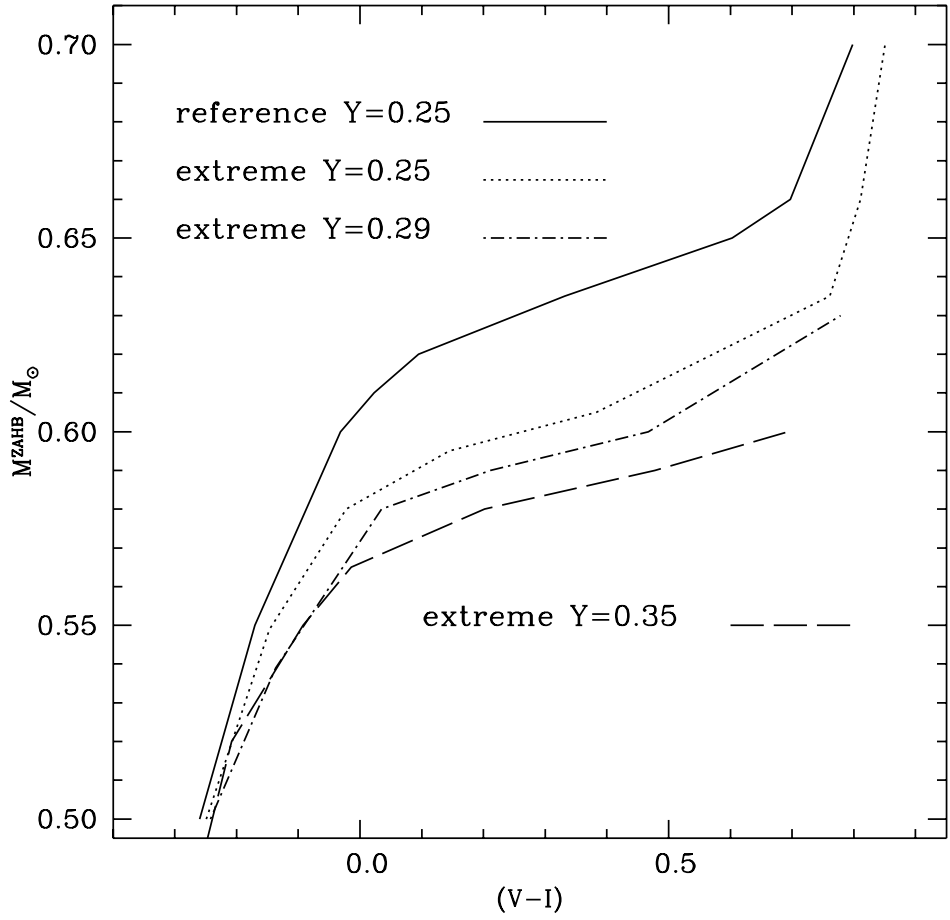}
\caption{Relationship between stellar mass and $(V-I)$ color for ZAHB models
with the labelled initial compositions.\label{fig6}}
\end{figure}

A general conclusion of this analysis is that ZAHB objects born out of the reference
and extreme (with $Y>$0.25) mixtures cannot coexist in comparable
quantities along the horizontal part of the HB. They
can coexist along the bluer HB sequence with the 
$Y=0.35$ population being eventually located at $(V-I)<-$0.15. 
As another possibility they can also coexist if objects 
with the reference composition occupy the $(V-I)\geq$0 region (and
eventually also the bluer part of the ZAHB) but the extreme
populations are located only at $(V-I)<$0.0. The opposite case looks less
feasible. In fact, if the mass loss is such that the $Y$=0.29 (or
$Y$=0.35) ZAHB objects   
populate the observed sequence at $(V-I)\geq$0.0, the reference ZAHB
will tend to be underluminous compared to the observed blue sequence;
this because an additional vertical shift of $\sim$ +0.2 or +0.5~mag has to be applied to
the reference and extreme ZAHBs displayed in Fig.~\ref{fig6}, 
in order for the extreme sequences to match the observed horizontal
part of the ZAHB. 

In a cluster like M13, whose HB is populated essentially at the blue
side, reference, $Y$=0.25 and 0.29 extreme populations can
coexist along the whole observed sequence; an eventual $Y=0.35$
subpopulation can coexist only at the bluer end of the observed HB.  
In case of M3, populated mainly in the horizontal part  
but also with a tail of objects at $(V-I)<$0.0, extreme (with $Y$=0.29
or $Y$=0.35) and reference subpopulations can coexist only if the
extreme components are located at the blue side of the observed HB. 

Figure~\ref{fig6} shows the stellar mass--$(V-I)$ relationship
along ZAHBs with the displayed 
four different chemical compositions. For a fixed value
of the mass, the ZAHB models with the anticorrelated mixture are
generally redder than the reference counterparts. This is due to the 
increased CNO abundance and lower He core mass. 
An increase of $Y$ at fixed metal mixture shifts the 
ZAHB location of a given mass slightly more to the red.
The initial mass of stars evolving at the RGB tip in the reference
isochrone (13~Gyr) and in the extreme isochrone with $Y$=0.25 (12.5~Gyr) is 
$0.8M_{\odot}$; for a 12.5~Gyr old
isochrone with the extreme mixture and $Y$=0.29, the initial value of
the evolving mass at the RGB tip is $0.75M_{\odot}$, while it is 
$0.67M_{\odot}$ when $Y$=0.35. These values, combined with the data in 
Fig.~\ref{fig6}, predict the minimum amount of mass lost along the
RGB, $\Delta M$, to be equal to $\sim$0.18~$M_{\odot}$ for the $Y$=0.29 isochrone, and 
$\sim$0.14~$M_{\odot}$ for the $Y$=0.35 isochrone, in order to
populate the ZAHB at $(V-I)<$0.0 and $(V-I)< -$0.15, respectively.

It may be interesting to analyze the situation if the Reimers~(1975) mass
loss law is applied. D'Antona et al.~(2002) have also discussed this
issue in detail. They assumed that cluster stars showing the CNONa anticorrelation
can be well described by isochrones with the same metal mixture of the
reference population and an increased initial He mass fraction
($Y$=0.29) and the same $Z$. Their results are now to a large 
extent validated by our analysis.
In their analysis they assumed standard values of the mean mass loss
along the RGB, obtained from the Reimers~(1975) law.  
The higher He isochrones will then be populated at the bluer part of the 
HB sequence, the reason for this being the
lower initial mass of the stars evolving at the RGB tip.  
As an example, we applied to our models a
Reimers~(1975) mass loss law with a standard $\eta$=0.3, and obtained 
an average $\Delta M~0.175 M_{\odot}$, that implies an actual mean value 
of the mass on the reference ZAHB equal to 0.625$M_{\odot}$. This
corresponds to a mean color $(V-I)\sim$0.2, in the horizontal part of the ZAHB. 
For a 12.5~Gyr old isochrone with the extreme mixture and $Y$=0.29, 
$\Delta M\sim 0.20M_{\odot}$ for the same $\eta$ (it is slightly
larger than for the reference isochrone). 
Under these conditions the corresponding ZAHB would be populated at a
mean color $(V-I)\sim -$0.1, given that the mean value of the mass along the ZAHB
is $\sim 0.55M_{\odot}$. Both reference and extreme
subpopulations satisfy one of the conditions described before for
their coexistence along the same HB sequence.


\section{Conclusions}

In this paper we present for the first time models, isochrones and
ZAHBs consistently computed for a metal mixture with the CNONa 
abundance anticorrelations observed in Galactic GCs.
Our analysis has shown that theoretical CMDs of clusters with abundance anomalies
produced by primordial pollution show a significant broadening only 
if the helium enhancement is extremely large ($Y$=0.35 in our 
study) in the stars showing anomalous abundances.
Also stellar luminosity functions from the MS to the RGB tip are
essentially unaffected. 
The spectroscopic results are therefore fully consistent with the
theoretical understanding of the photometry. 
From the photometry alone one cannot decide whether 
the AGB-pollution (increased helium) or the RGB-scenario 
(normal helium) is more likely. Detailed quantitative analyses 
of the luminosity function in the bump region might give some clues, 
but only if the fraction of objects with extreme values of the
anticorrelations is high. In this case relative shifts of the bump central
magnitude of the order of 0.05-0.10 magnitudes are possible, depending on
the amount of initial He in the extreme population and its relative
weight compared to the reference component. This kind of relative 
differences are however of the same order of magnitude or smaller than 
the accuracies of most empirical determinations of the bump brightness
(see, e.g., Caloi \& D'Antona~2005). 
It is however intriguing to notice that Caloi \& D'Antona~(2005) found a
variation of 0.14$\pm$0.09 in the quantity $\Delta
V_{TO}^{bump}=V_{TO}-V_{bump}$ between M3 and M13 (the latter showing
the larger value). Assuming the same age and the 
same [Fe/H] for the two clusters 
(see, e.g., the discussion in Caloi \& D'Antona~2005)  
our Monte-Carlo simulations of a composite 50:50 reference-extreme population discussed
above predict an increase of $\Delta V_{TO}^{bump}$ by 0.11~mag (we
considered the magnitude of the center of the bump region) when
the initial He abundance of the extreme population increases from $Y$=0.25 to 
$Y$=0.29, and by 0.20~mag, when it goes from $Y$=0.25 to 
$Y$=0.35. These differences are due to the change of the bump
magnitude with changing $Y$ of the extreme population (brighter
for increasing $Y$) but also to the change of the TO brightness of
the combined population (fainter for increasing $Y$ of the extreme
population, as derived from our composite populations by  
determining the color distribution of the synthetic MS stars 
as a function of $M_V$). 
A difference in the initial $Y$ abundance for the extreme
components of these two clusters (higher $Y$ for M13, that also shows 
the bluer HB) could explain the difference of the 
observed $\Delta V_{TO}^{bump}$ values, as
also suggested by Caloi \& D'Antona~(2005).

We have also studied the distribution of stars along the ZAHB and --
under most general assumptions about the mass lost long the RGB phase --  
derived constraints about the relative location of
objects with and without abundance
anomalies, along the observed horizontal branches of globular clusters.
Spectroscopic identification of HB objects with abundance anomalies
will be decisive in determining the amount of mass lost by their RGB progenitors, 
on the basis of the location along the HB sequence. Spectroscopy of
HB stars might possibly also give some indirect hint of the initial 
He abundance (hence whether the source
of primordial anomalies is pollution from AGB or RGB objects) in stars
showing the chemical anomalies. As an example, if sizable samples of 
HB stars with both reference and extreme metal distribution are coexisting along the
horizontal part of the HB of a cluster like M3, this would favor -- on
the basis of the discussion in the previous section -- a 'normal'
initial He abundance in the stars with extreme composition, 
hence pollution from RGB stars.

We close with a final comment about the effect of the MgAl anomalies. 
In the our analysis we have neglected the effect of the MgAl
anticorrelation (the Al abundance is higher when the Mg content 
is lower and the Na abundance is higher) observed in
several clusters. Its inclusion, however, would not have altered 
our main results, for the following reasons. Both Al 
(that is increased at most by
$\sim$1.0~dex) and Mg (that is decreased by at most 
$\sim$0.4~dex) do not contribute appreciably to
the energy generation, therefore their effect on evolutionary
timescales and luminosities is negligible. Along the MS and TO phases 
a change of Mg and Al does not alter the 
evolution through opacity effects either, because they do not contribute
substantially to the Rosseland opacities in the temperature range that affects
the MS location (e.g., Salaris, Chieffi \& Straniero~1993). The He core
masses along the RGB evolution are also unaffected (hence also the
ZAHB brightness will stay unchanged). 
As for the RGB color, it is completely determined by the low temperature
opacities (e.g., Salaris et al.~1993); as a test we have computed 
low temperature opacities for a metal mixture with Mg decreased by 
0.4~dex and Al increased by 1.0~dex with
respect to a reference [Fe/H]$\sim -$1.3 scaled solar metal distribution. 
Comparisons in the relevant density-temperature range for the appropriate
hydrogen abundance, show that at the same Fe abundances the two sets of
opacities show negligible differences, of less than 1\%. This means
that the MgAl anticorrelation leaves also the RGB location unaffected.

\acknowledgments
We thank our anonymous referee for comments that helped us 
to improve the presentation of our results.

\clearpage



\clearpage

\end{document}